\documentclass[sigconf, authorversion]{acmart}

\usepackage{lipsum}
\usepackage{enumitem}
\usepackage{xurl}

\copyrightyear{2021} 
\acmYear{2021} 
\setcopyright{acmlicensed}\acmConference[EPIQ'21]{Workshop on the Evolution, Performance and Interoperability of QUIC}{December 7, 2021}{Virtual Event, Germany}
\acmBooktitle{Workshop on the Evolution, Performance and Interoperability of QUIC (EPIQ'21), December 7, 2021, Virtual Event, Germany}
\acmPrice{15.00}
\acmDOI{10.1145/3488660.3493804}
\acmISBN{978-1-4503-9135-1/21/12}

\usepackage{hyperref}
\usepackage{units}

\usepackage{algorithmic}
\usepackage{algorithm}

\usepackage{multirow}

\usepackage{acro}
\makeatletter
\@ifpackagelater{acro}{2017/02/17}{\@ifpackagelater{acro}{2019/09/23}{}{
    \let\IDeclareAcronym\DeclareAcronym
    \renewcommand{\DeclareAcronym}[2]{%
        \IDeclareAcronym{#1}{%
        #2,foreign-plural={}
        }
    }
}}{}

\DeclareAcronym{RTT}{
  short        = RTT,
  long         = round-trip time
}

\DeclareAcronym{VEC}{
  short        = VEC,
  long         = Valid Edge Counter
}

\DeclareAcronym{EFM}{
  short        = EFM,
  long         = Explicit Flow Measurements
}

\DeclareAcronym{MAU}{
  short        = MAU,
  long         = Match-Action Unit
}

\DeclareAcronym{ALU}{
  short        = ALU,
  long         = arithmetic logic unit
}

\DeclareAcronym{CID}{
  short        = CID,
  long         = connection ID
}

\newcommand{\afblock}[1]{\noindent{\textbf{#1}}}
\newcommand{\takeaway}[1]{\noindent{\textbf{Takeaway.}} \textit{#1}}

\usepackage{prettyref}
\newrefformat{chap}{Chapter~\ref{#1}}
\newrefformat{sec}{Sec.\,\ref{#1}}
\newrefformat{sub}{Sec.\,\ref{#1}}
\newrefformat{subsec}{Sec.\,\ref{#1}}
\newrefformat{eq}{Eq.~(\ref{#1})}
\newrefformat{fig}{Fig.~\ref{#1}}
\newrefformat{plot}{Fig.~\ref{#1}}
\newrefformat{ex}{Example~\ref{#1}}
\newrefformat{list}{List~\ref{#1}}
\newrefformat{tab}{Table~\ref{#1}}
\newrefformat{enum}{Case~(\ref{#1}.)}
\newrefformat{def}{Definition~\ref{#1}}
\newcommand{\pref}[1]{\prettyref{#1}}

\usepackage{circledsteps}

\usepackage{tikz}

\usepackage[caption=false,font=footnotesize]{subfig}

\usepackage{lipsum}

\begin{document}

\title{Tracking the QUIC Spin Bit on Tofino}

\author{Ike Kunze}
\email{kunze@comsys.rwth-aachen.de}
\orcid{0000-0001-8609-800X}
\affiliation{%
  \institution{RWTH Aachen University}
  \country{Germany}
}

\author{Constantin Sander}
\email{sander@comsys.rwth-aachen.de}
\affiliation{%
  \institution{RWTH Aachen University}
  \country{Germany}
}

\author{Klaus Wehrle}
\email{wehrle@comsys.rwth-aachen.de}
\orcid{0000-0001-7252-4186}
\affiliation{%
  \institution{RWTH Aachen University}
  \country{Germany}
}

\author{Jan R\"uth}
\email{rueth@comsys.rwth-aachen.de}
\orcid{0000-0002-4993-3210}
\affiliation{%
  \institution{RWTH Aachen University}
  \country{Germany}
}

\renewcommand{\shortauthors}{Kunze et al.}

\begin{abstract}
QUIC offers security and privacy for modern web traffic by closely integrating encryption into its transport functionality.
In this process, it hides transport layer information often used for network monitoring, thus obsoleting traditional measurement concepts.
To still enable passive RTT estimations, QUIC introduces a dedicated measurement bit -- the \emph{spin bit}.
While simple in its design, tracking the spin bit at line-rate can become challenging for software-based solutions. 
Dedicated hardware trackers are also unsuitable as the spin bit is not invariant and can change in the future.

Thus, this paper investigates whether P4-programmable hardware, such as the Intel Tofino, can effectively track the spin bit at line-rate.
We find that the core functionality of the spin bit can be realized easily, and our prototype has an accuracy close to software-based trackers.
Our prototype further protects against faulty measurements caused by reordering and prepares the data according to the needs of network operators, e.g., by classifying samples into pre-defined RTT classes.
Still, distinct concepts in QUIC, such as its connection ID, are challenging with current hardware capabilities.
\end{abstract}



\maketitle

\section{Introduction} 
\label{sec:introduction}

Network monitoring is essential for managing modern networks~\cite{juniper:2017:selfdriving}, and current approaches range from embedding measurement data into production traffic~\cite{ben-basast:sigcomm2020:pint,p4lang:github:int,huang:SIGCOMM2020:omnimon,zhou:sigcomm2020:netseer,RFC8321} to passively observing the traffic~\cite{cociglio:AcmToN2019:multipoint}.
The latter option is generally seen as the go-to choice for large-scale access networks as it entails a low overhead. 
Corresponding techniques typically exploit the wire image~\cite{RFC8546} of production traffic to derive network KPIs providing further insights, e.g., w.r.t.\ SLAs, abnormal latencies, or required topology changes.

To cope with ever-increasing network sizes and speeds, network monitoring and telemetry tasks are increasingly performed directly on the data plane of programmable hardware switches~\cite{hauser:arxiv2021:p4SurveyTuebingen,kfoury:ieeeaccess2021:theOtherP4Survey,manzanares:ieeeaccess2021:passiveInBandTelemetry}.
However, these platforms come with specific constraints 
that render implementations not as straightforward as conventional software solutions~\cite{kunze:IM2021:AQMonTofino}.
Still, related work has recently shown that traditional TCP RTT estimations leveraging sequence numbers (SEQs) and ACKs~\cite{RFC8321,benko:globecom2002:tcppacketloss,RFC8546} can be efficiently realized on such devices~\cite{chen:spin2020:RTTmeasurements}, thus enabling measurements at line-rate without deploying dedicated measurement hardware or the need for challenging software-based processing of large quantities of packets.

In contrast to TCP, QUIC has a drastically reduced wire image as it encrypts its signaling information, including its notion of SEQs and ACKs, which prevents conventional SEQ/ACK-based measurements.
Recognizing this challenge, the IETF standardized a dedicated measuring bit within QUIC -- the \emph{spin bit}~\cite{RFC9000} -- which is exempted from encryption and allows a passive observer to estimate RTTs.
Tracking the spin bit is relatively easy in software, e.g., using Spindump~\cite{ericsson:2021:spindump}, which can further compensate for errors caused by reordering~\cite{devaere:2018:imc:threebits}.
However, as with TCP, ever-increasing data rates make hardware implementations necessary, but purpose-built hardware comes with additional cost and the spin bit is an optional, non version-independent QUIC feature, i.e., hardware investment now may soon prove to be wasted should future QUIC versions modify the spin bit.
Thus, re-programmable hardware seems like a promising solution to implement the conceptually simple spin bit mechanism.

This paper thus explores how the spin bit mechanism maps to a commodity programmable switch, the Intel Tofino.
We identify several challenges raised by the combined characteristics of programmable data planes, QUIC, and the spin bit and discuss and present potential solutions.
Specifically, we start with a naïve spin bit tracking mechanism that we complement with several additional processing steps.
In particular, we test two mechanisms for protecting against reordered spin bit flanks and explore different means to export the measurements from the data plane: \emph{periodically} and \emph{event-driven}.
We further approximate RTT averages and show how to utilize operator-defined RTT classes to classify measurements already on the data plane.
Our evaluations compare the accuracy of RTT estimations by our P4-based spin bit tracker~\cite{kunze:2021:spin-tracker} to ground-truth estimates provided by Spindump, revealing that a pure spin bit implementation is possible on Tofino.
Similarly, our more evolved concepts prove to be effective, although at the cost of higher resource consumption. 

\afblock{Structure.}
In \pref{sub:p4_programmable_networking_hardware}, we first concisely introduce fundamental aspects of programmable networking hardware.
\pref{sec:tracking_the_spin_bit_on_the_dataplane} then presents the design and implementation of our P4-based spin bit tracker, whose performance we evaluate in \pref{sec:evaluation}.
Finally, we discuss related work (\pref{sec:related_work}) and limitations (\pref{sec:limitations}) of our tracker before concluding the paper.

\section{P4-Programmable Hardware} 
\label{sub:p4_programmable_networking_hardware}
The current generation of programmable networking hardware is usually split into an ASIC-based data plane that processes packets at line-rate and a slower CPU-based control plane that configures and manages said data plane.
The data plane follows a pipeline-based model, which allows for a certain degree of programmability, e.g., using P4~\cite{bosshart:ccr2014:p4}. 

Enclosed by programmable parsers and de-parsers that first extract packet information and later re-assemble the packets, the heart of the pipeline consists of a fixed number of configurable stages, each containing so-called \acp{MAU} to which all desired computations are mapped.
More specifically, the \acp{MAU} define \emph{matching} logic on SRAM or TCAM to perform key-based lookups while the \emph{action} part allows for a small number of simple arithmetic and logical operations.
One current state-of-the-art programmable switch, the Intel Tofino, e.g., enables multiplications involving powers of two with statically defined multiplicands.

So-called \emph{registers} are memory constructs that can be read and written from within the data plane.
However, due to the pipeline model, these registers and all other memory can only be accessed within the associated \acp{MAU}, i.e., once in the pipeline (\emph{single-access rule}), which considerably limits read-update-write concepts on state variables.
Tofino enables read-update-write functionality in the form of small microprograms.
Overall, the nature of these pipelines that are designed for very high-speed processing of packets with latencies that are both low and fixed makes programming, in comparison to much more flexible but slow CPU-based processing, more difficult.
In the following, we present how we still implement a basic spin bit tracking mechanism as well as additional heuristics on Tofino. 

\begin{figure}[t]
  \centering
  \includegraphics[width=0.95\columnwidth]{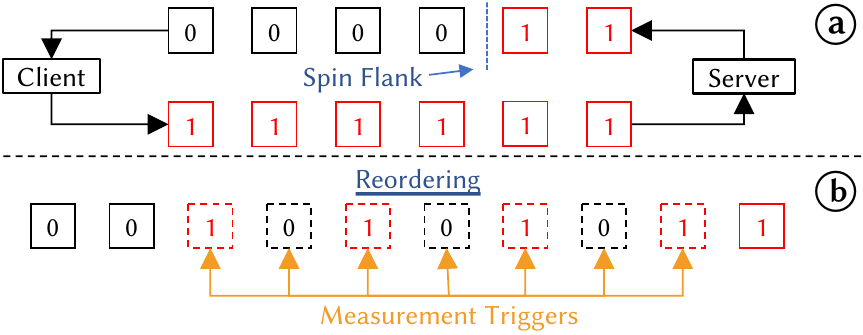}
  \caption{\Circled{a}: The client of a connection always \emph{flips} the spin bit; the server \emph{reflects} it. \Circled{b}: Reordering near the spin signal flanks can cause faulty measurements.}
  \label{fig:spinbit}
\end{figure}


\section{Tracking The Spin Bit On Tofino} 
\label{sec:tracking_the_spin_bit_on_the_dataplane}
The spin bit is QUIC's answer to the call for explicit measurement support by Allman et al.~\cite{allman:2017:ccr:measurability} and exposes a measurable square wave signal onto a QUIC flow enabling RTT measurements.
For this, the client always \emph{flips} the spin bit it received while the server \emph{reflects} it, as shown in \pref{fig:spinbit} \Circled{a}.
An observer can then determine the flow's RTT by measuring the time between consecutive spin bit flips.

\begin{figure*}[t]
  \centering
  \includegraphics{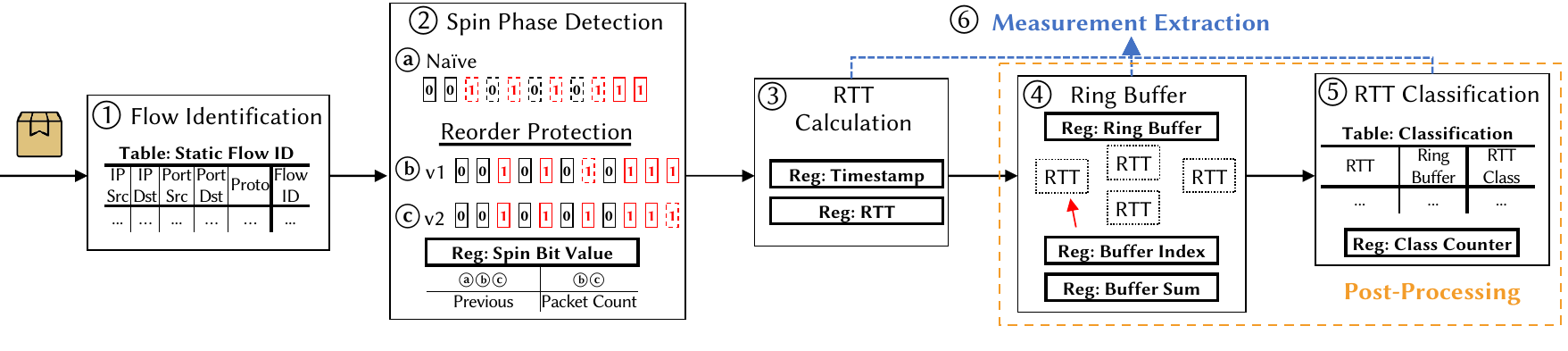}
  \caption{A five-step spin bit tracking pipeline. After an initial flow identification (\Circled{1}) and a spin phase detection (\Circled{2}), our pipeline calculates the current RTT (\Circled{3}). It further keeps track of the last $N$ measurements in a ring buffer (\Circled{4}) and can classify the current RTT based on those last measurements (\Circled{5}).}
  \label{fig:pipeline}
\end{figure*}

Based on this simple mechanism, we have implemented a P4-based spin bit tracker~\cite{kunze:2021:spin-tracker} in the Ingress of an Intel Tofino switch.
\pref{fig:pipeline} illustrates our overall pipeline, which centers around the core spin bit tracking (\Circled{2}) and RTT calculation (\Circled{3}) mechanisms.
In particular, we first have to \emph{identify a flow} as the spin bit is always set on a per-flow basis (\Circled{1}).
We then study three variants to detect a spin phase-change, mainly owed to the spin bit's susceptibility to reordering (\Circled{2}).
After computing the RTT (\Circled{3}), we investigate how additional post-processing concepts found in software-based observers, such as averaging (\Circled{4}) or range-based filtering (\Circled{5}), can be realized on programmable switches to reduce the amount of information that needs to be communicated to the network operator.
Finally, we explore different ways of extracting the measured RTTs from the data plane (\Circled{6}).

In the following, we describe the design of the different components and the challenges we faced when realizing them, starting with the flow identification.

\subsection{Flow Identification -- \Circled{1}} 
\label{sub:flow_identification}
The spin bit is a per-flow signal, i.e., we must be able to differentiate between flows.
QUIC offers variable-length \acp{CID} that identify flows even when IPs change (e.g., on NAT rebinding).
The CID is negotiated in the handshake using long header packets that denote the sender's and receiver's \ac{CID}, each preceded by its respective length~\cite{RFC9000}.

While the \ac{CID} is not encrypted, using it for identifying flows on a switch is still challenging.
We first require some mapping to \ac{CID} length to extract the \ac{CID} from short header packets (containing the spin bit) as this header, unfortunately, lacks a length field~\cite{RFC9000}.
This mapping can either be manually configured or extracted from long header packets during the handshake.
Since a server may change the initially client-chosen destination \ac{CID}, one needs to update subsequently observed \ac{CID} length fields.
We prototyped the \ac{CID} length extraction from long header packets to show its feasibility but are currently not using it and instead set up the mapping manually.
However, having two variable-length fields (20 byte max.\ length.) is too taxing on the Tofino's deparser, yet, representing each possible length by one specific field does work, although this has a heavy cost on the deparser, too. 

For the actual \ac{CID} extraction in short header packets, we can speculatively parse 20 bytes of \ac{CID}.
We then match on the five-tuple\footnote{Depending on mobility or NAT rebinding at which side, just one endpoint's IP and port may be more suitable.} to find the length of the \ac{CID}.
If the length is non-zero, we extract the corresponding subset from the 20 bytes as the actual \ac{CID}.
The \ac{CID} (or the five-tuple for zero-length \acp{CID}) is then hashed as a flow ID that identifies our spin bit measurements.
While we did implement this on Tofino, the sequential dependencies, together with our spin bit measurements and post-processing steps, in sum, exhaust the available stages.
Our prototype currently focuses on the measurement-related capabilities and maps flows only by five-tuple as depicted in \pref{fig:pipeline}--\emph{Table:\,Static Flow ID}.

Next, after identifying a flow, the spin bit tracker has to correctly detect spin bit phase transitions.

\subsection{Spin Phase Detection -- \Circled{2}} 
\label{sub:spin_phase_detection}
The spin bit is known to be susceptible to packet reordering~\cite{devaere:2018:imc:threebits} as out-of-order packets near the flanks of the square wave signal will cause invalid RTT readings.
For example, in \pref{fig:spinbit} \Circled{b}, each packet within the affected spin flank (dotted packets) produces faulty estimates. 
Simply triggering a new measurement once the spin bit changes as done in our naïve variant (\Circled{a}) might thus lead to invalid results in times of reordering.
While reorder protections can already detect and discard many erroneous measurements~\cite{devaere:2018:imc:threebits}, they rely on performing the measurement first, then applying heuristics, and finally rejecting invalid measurements.
Such concepts are not directly possible in the data plane due to the single-access rule.
While workarounds using packet recirculation are possible, they add non-negligible overhead.
We thus experiment with two approaches to actively protect against reordering.
They are inspired by the reorder protection proposed for the square bit, an additional measurement extension that enables packet loss measurements~\cite{ietf-draft-explicit-flow-measurements}.

\afblock{Reorder Protection v1 -- \Circled{b}.}
Our first approach directly follows the protection of the square bit.
A phase transition, and thus a measurement, is only triggered as soon as $N$ packets of the new flank have been registered, e.g., after the third 1-bit in \pref{fig:pipeline} for $N=3$.
However, in scenarios with significant reordering around a flank or if the spin bit is greased, i.e., set to arbitrary values when the spin bit functionality is not enabled, this simple form of protection might still yield many invalid measurements.

\afblock{Reorder Protection v2 -- \Circled{c}.}
Our second approach tries to resolve these shortcomings by requiring the same number of packets but in direct succession, putting a stronger requirement on the phase-change.
Depending on the threshold value, this scheme should filter greased spin bits and also most invalid measurements in case of actual reordering.

\afblock{Implementation.}
We implement each of the three phase-change detection approaches using a single register that tracks the current value of the spin bit (\pref{fig:pipeline}--\emph{Reg:Spin Bit Value}).
However, the reorder protection variants require more memory as they do not only track the spin bit (left) but also the number of seen packets (right).
In our prototype, we deploy all three concepts in parallel to allow for easy switching between the different modes.
Upon detecting a phase-change, the subsequent RTT calculation is triggered.

\subsection{RTT Calculation -- \Circled{3}} 
\label{sub:rtt_calculation}
We calculate the RTT using two registers, one tracking the timestamp of the previous spin bit flip (\pref{fig:pipeline}--\emph{Reg:\,Timestamp}), the other storing the calculated RTT (\pref{fig:pipeline}--\emph{Reg:\,RTT}).
For measuring time, Tofino provides a 48-bit timestamp with nanosecond resolution.
For simplicity and to save resources, our current spin bit tracker prototype uses a well-chosen 16-bit slice of this switch-local timestamp to derive a timestamp with near-millisecond resolution (one timestep $\approx$ 1.049 ms) that wraps around after roughly 69 seconds; we plan to evaluate the suitability of the chosen slice in future work (see \pref{sec:limitations}).
Using the previously described mechanism, this core component of the spin bit measurement functionality nicely maps to Tofino.
In the following, we show how these simple measurements can be leveraged for additional post-processing steps, such as determining average RTTs or classifying measurements.

\subsection{Post-Processing} 
\label{sub:post_processing}
Network operators might not be interested in the exact RTTs of specific flows but rather in a rough distribution of the occurring RTTs or whether the RTT of flows fluctuates.
Consequently, there is potential in early post-processing on the data plane to reduce the amount of information that needs to be communicated between the data and the control plane.

\afblock{RTT Averaging (Ring Buffer) -- \Circled{4}.}
For classification tasks, it is essential to maintain statistics on the already seen RTT values.
For instance, software-based RTT tracking tools, such as Spindump, use a moving average of the current values.
However, such techniques are not easy to implement on the data plane, as, e.g., required divisions are typically not supported.
To still capture some form of average RTT, we implement a simple ring buffer on the data plane that stores the last $N$ RTT values of a flow (\pref{fig:pipeline}--\emph{Reg:\,Ring Buffer}).
We further store the sum over all ring buffer entries for a particular flow in another register and update it whenever we exchange entries of the ring buffer (\pref{fig:pipeline}--\emph{Reg:\,Buffer Sum}).
A third register tracks the read/write position of the ring buffer (\pref{fig:pipeline}--\emph{Reg:\,Buffer Index}).
Knowing $N$, the control plane can calculate the mean RTT of a flow without having to react on every new result or to keep state.

\afblock{RTT classification -- \Circled{5}.}
While explicit information on the current RTTs is important, we believe that network operators are more interested in the general traffic characteristics and the stability of their network's performance.
Consequently, we propose deploying a simple form of RTT classification already in the data plane that builds upon our ring buffer solution.
In our concept, the control plane can configure different RTT ranges, and the data plane then accurately tracks which of these ranges are hit by the flows currently under study.
This information could, e.g., be used to create RTT histograms.
Our prototype uses a simple range setting with three classes, one designed to capture greased spin bits and reorderings ($<5$ms), one intended to capture stable measurements ($+/- 10\%\ mean RTT$), and one containing the regions in between and above, thus capturing unstable measurements.
We leave finding an optimal configuration for these RTT classes to future work.
The RTT classification deploys one \ac{MAU} using range matches of the current RTT and the ring buffer RTT to classify the current RTT (\pref{fig:pipeline}--\emph{Table: Classification}).
An additional register then tracks how often each of the classes has been hit (\pref{fig:pipeline}--\emph{Reg:\,Class Counter}).

\begin{figure}[t]
  \centering
  \includegraphics{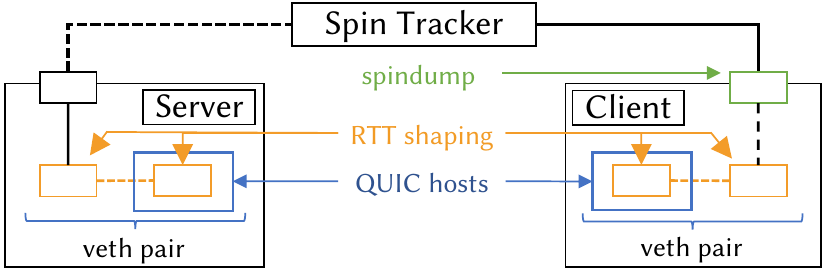}
  \caption{One Tofino switch running our spin tracker interconnects two QUIC end-hosts.}
  \label{fig:testbed}
\end{figure}

\subsection{Measurement Report Intervals -- \Circled{6}} 
\label{sub:measurement_report_intervals}
In contrast to TCP measurements, there is at most one measurement per RTT for the spin bit.
Thus, there are different options for notifying the control plane of new measurements.

One option is leveraging the switch-local control plane to read out the data plane registers in fixed intervals (\emph{periodic}).
While this allows for well-controlled read intervals, it might be challenging to find the best readout interval for each situation that ensures a truthful representation of the signal.

As a more dynamic alternative, we let the data plane send out one dedicated packet containing the latest measurement results whenever a new spin cycle has been detected (\emph{event-based}).
For this, we clone the packet that triggered the measurement and then modify the clone to contain the RTT estimations before redirecting it to the control plane.
This approach is more robust to changing flow dynamics yet requires additional data-to-control plane bandwidth.

\section{Evaluation} 
\label{sec:evaluation}
We evaluate the accuracy and capabilities of our prototype in a testbed where two end-hosts are interconnected by an Intel Tofino switch running the spin bit tracker, as illustrated in \pref{fig:testbed}.
We shape delay using \emph{tc netem} on the interconnecting links of the veth pairs on each end-host and configure a steady bandwidth of \unit[10]{Mbps}\footnote{This choice is sufficient for our evaluation as our prototype does not use recirculation. The measurements thus scale with the RTT and not with the bandwidth.} on the client's ingress interface.
For measuring RTTs, we create QUIC traffic using aioquic~\cite{laine:2021:aioquic} at the QUIC hosts (blue) and capture the output of our tracker, the qlog~\cite{marx:anrw2020:qlog} information of aioquic, and the traffic at the client's ingress interface, which we subsequently analyze using Spindump~\cite{ericsson:2021:spindump}.
While the Spindump results provide a groundtruth regarding the accuracy of plain spin bit measurements, the qlog data serves as a groundtruth for the actual RTT.
To accurately compare the timestamps of the different involved entities, we use PTP~\cite{IEEE1588} to synchronize the clocks before each measurement run.
For each setting, we perform 30 measurement runs and report mean values with 99\% confidence intervals.

\subsection{Web Traffic Performance} 
\label{sub:web_traffic_performance}
\begin{figure}[t]
  \centering
  \includegraphics{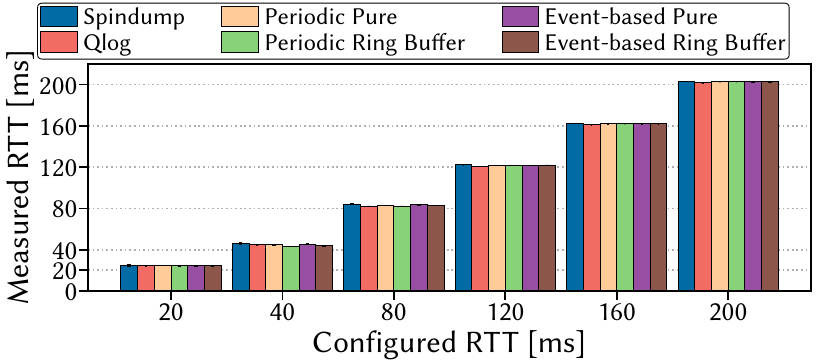}
  \caption{Measured mean RTTs for different configured RTTs and a download of \unit[2]{MB}.}
  \label{plot:averageRTT}
\end{figure}

We first investigate how well our spin tracker can track the RTT of standard web traffic.
For this, we configure different RTTs and let our client request different-sized files using HTTP/3 from the server.
\pref{plot:averageRTT} shows the measured mean RTT across all measurements when downloading a \unit[2]{MB} file.

\begin{figure*}
\centering
\subfloat[Measured mean RTTs]{\label{plot:averageRTTReorder}\includegraphics{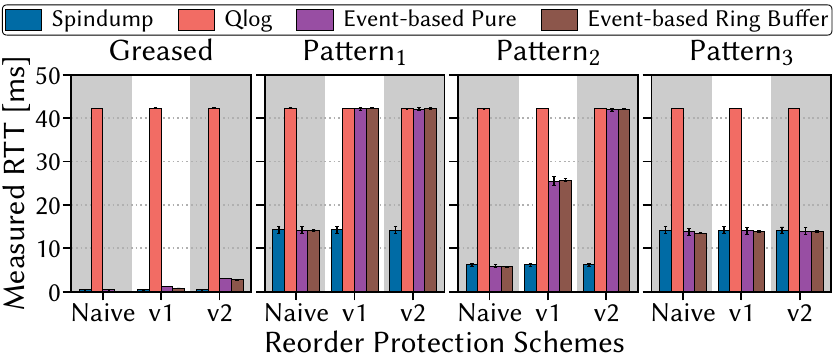}}\hfill
\subfloat[Mean RTT classification counter values]{\label{plot:classificationHistogram}\includegraphics{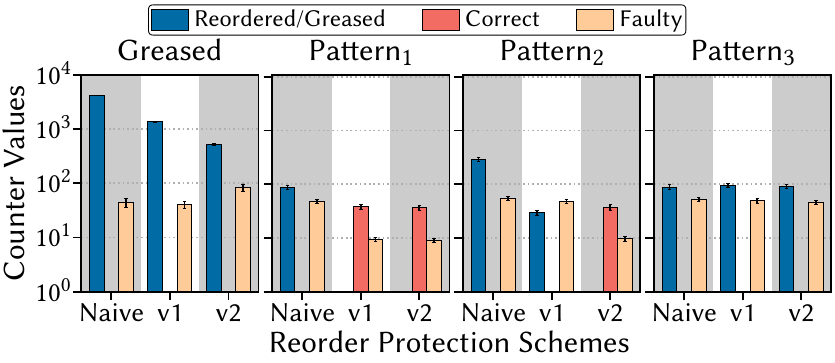}}
\caption{Results for four spin bit patterns, an RTT of \unit[40]{ms}, and a download of \unit[10]{MByte}.}
\label{plot:reorderStuff}
\end{figure*}

The individual (\emph{pure}) readings of the \emph{periodic} and \emph{event-based} solutions are very~close to the \emph{Qlog} and \emph{Spindump} estimates, showing the general feasibility of performing spin bit measurements on the data plane.
The \emph{ring buffer} shows steady results, proving that RTT approximations using a ring buffer on the data plane with subsequent average calculations on the control plane are feasible.

However, when experimenting with different bottleneck queue sizes and bandwidths (not shown), we found that the oversampling of our interval-based variant (\unit[5]{ms} readout interval) can lead to an RTT underestimation when small queues build up.
In particular, the induced queuing delay is only eventually picked up by the spin bit, while the interval-based readout oversamples readings in the transition periods, thus causing an underestimation of the RTT.
Consequently, selecting suitable readout intervals, especially for multi-flow settings with different latencies, is essential and non-trivial as too high intervals will cause measurements to be missed.
At the same time, too short intervals might oversample the available measurements and cause an untruthful representation of the RTTs.

\subsection{Reorder\,Protection \& RTT\,Classification} 
\label{sub:reorder_detection}
We next study how we can use the reorder protection mechanisms and the RTT classification to account for greased spin bits and packet reordering occurring at the spin bit flanks, both causing invalid measurements.
To focus our evaluation, we artificially create four distinct traffic patterns targeting a reordering threshold of three packets.
We leave investigating an optimal sizing of the reordering threshold for future work.  

\afblock{Spin Bit Patterns.}
\emph{Greased} sets the spin bit randomly to create a random spin bit sequence.
In contrast, three reordering patterns produce fixed sequences.
\emph{Pattern$_{1}$} inserts one duplicate packet of the old flank after an actual spin flank which causes two invalid readings in short succession.
\emph{Pattern$_{2}$} repeats this process for the first three packets of the new flank, thus triggering six invalid measurements to circumvent reorder protection v1.
Finally, \emph{Pattern$_{3}$} transmits a sequence of three packets of the old flank as soon as there were three packets of the new flank, thus disabling reorder protection v2.
The first two patterns represent random reordering, while the latter could, e.g., be caused by faulty load balancing. 

\afblock{Results.}
\pref{plot:reorderStuff} shows (a) the measured mean RTT of our event-based approach (\emph{pure} and \emph{ring buffer}) in comparison to end-host estimates (\emph{Qlog}) and \emph{Spindump}, as well as (b) the mean counter values of our RTT classification tables, denoting how often a class was hit, in a setting with an RTT of \unit[40]{ms} and a download size of \unit[10]{MB} when inducing the four spin bit patterns.

The end-hosts automatically filter the reordering and produce realistic latency estimations in all settings.
In contrast, when not using reorder protection, the Spindump ground truth and our spin tracker expectedly underestimate the actual RTT for all patterns.
However, our RTT classification is able to classify these measurements as faulty, which can be seen in the high values of the corresponding counters (reordered/greased and faulty).
No readings are classified as correct because this requires a stable RTT average which is not the case for these frequent reorderings.

Reorder protection v1 effectively protects against \emph{Pattern$_{1}$}.
It further yields more reasonable results for \emph{Pattern$_{2}$} compared to no protection as it filters many invalid edges and only produces one invalid reading for each spin flank.
Consequently, its RTT estimates are roughly half of the real RTT.

Reorder protection v2 protects against \emph{Pattern$_{1}$} and \emph{Pattern$_{2}$} as it puts the tightest requirements on a flank.
Most measurements are classified as \emph{stable}, mixed with a few faulty readings, which can be explained by the queue build-up already noted in \pref{sub:web_traffic_performance}.
Yet, v2 does not cope with \emph{Pattern$_{3}$}, i.e., sequences of reordered packets exceeding its threshold.

Finally, none of our schemes fully protect against greasing, although they filter many invalid readings and detect all faulty measurements accordingly. 

\takeaway{Our results show that it is possible to i) filter invalid edges from the outset or ii) at least identify invalid/unstable measurements already on the data plane, thus showing potential for post-processing according to the needs of network operators.
Depending on the expected level of reordering, our proposed protection schemes can be configured with different thresholds for tailored protection.
Similarly, the RTT classes are configurable to the expected latency ranges and offer a very focused view of the network characteristics.}

To achieve optimal cost-efficiency, our approach should be deployed on hardware already used for standard networking functionality.
We thus finally analyze the resource use of our prototype.

\subsection{Resource Footprint} 
\label{sub:resource_footprint}
To judge the practicability of our approach, we finally dissect the relative resource use of the different components of our prototype in a configuration that is able to track up to 10~000 flows with a ring buffer size of $4$ and $8$ different RTT classes.   

As can be seen in \pref{tab:resource-footprint}, a \textbf{base} implementation of our pipeline (Steps \Circled{1} -- \Circled{3}~\Circled{a}) together with a simple L2/L3 forwarding logic is very light on resources and leaves plenty of room for other functionality.
The reorder protection (\Circled{3} \Circled{b}/\Circled{c}) also comes with minor additional resource use, even when simultaneously adding both variants.
In contrast, the ring buffer mechanism (\Circled{4}) requires more resources and additional pipeline stages.
Still, even when deploying the RTT classification (\Circled{5}) or all components simultaneously (\textbf{Full}), the overall resource use is still modest, with ALUs being the most used component with 13.5\% of the available resources.
While our full spin bit tracker requires resources at all available stages due to sequential dependencies, its modest resource use shows the practicability of deploying a spin bit tracker on programmable networking hardware alongside other networking functionality.
However, adding a \ac{CID}-based flow identification overstretches the sequential capabilities, illustrating that the available resources can make prototypes fail even though they are conceptually possible. 

\begin{table}[t]
  \centering
    \caption{Resource footprint of different compositions of our spin tracker in \% of the available resources.}
  \label{tab:resource-footprint}
  \def\arraystretch{0.5}
  
  \begin{tabular}{l|r|r|r|r|r}
    \toprule

  \textbf{Resource [\%]} & \textbf{Base} & +\Circled{3} \Circled{b}/\Circled{c} & +\Circled{4} & +\Circled{4}\Circled{5} & \textbf{Full} \\
    \midrule
    Stages & 41.7 & 41.7 & 58.3 & 100     & 100\\

    SRAM & 3.4     & 4.2      & 5.2   & 6.0   & 6.7\\

    TCAM & 0.0     & 0.0 & 0.0   & 3.5     & 3.5\\

    Hash Bits & 4.6 &        5.2 &     5.8 &       6.6    & 6.8 \\

    VLIW Actions & 2.3     & 3.1   & 3.6     & 5.2    & 6.0 \\

    Match X-bars & 2.4               & 2.5                 & 3.2               & 3.8    & 3.7 \\

    ALUs & 6.23                 & 8.3             & 10.4            & 11.5     & 13.5 \\
        \bottomrule
  \end{tabular}
\end{table}


\section{Related Work} 
\label{sec:related_work}
Tracking flow latencies on the data plane has already been studied for TCP.
Ghasemi et al.~\cite{ghasemi:sosr2017:dapper} propose Dapper to track various TCP metrics, including the RTT, which they measure using the SEQ/ACK methodology.
In contrast, Chen et al.~\cite{chen:spin2020:RTTmeasurements} completely focus on RTT measurements and show that these can be conducted using a Tofino switch.
They first store timestamps of outgoing packets in a multi-stage hash table and then match those to timestamps of incoming packets, cross-relating the packets using hashes of the five-tuple.
Apostolaki et al.~\cite{apostolaki:sosr2021:routescout} perform RTT measurements based on the TCP handshake leveraging SYN/SYNACK packets which they map using a Counting Bloom Filter.
A similar methodology could also be applied to measure the RTT of QUIC's handshake.   

\section{Discussion} 
\label{sec:limitations}

\afblock{Flow Identification.}
While QUIC has an explicit \ac{CID}, its lack of a length indicator in short header packets complicates a \ac{CID}-based flow tracking as the connection establishment must be observed.
Even though possible to implement, it i) involves many sequential steps that challenge a placement with further functionality on a fixed number of stages, and ii), depending on what scenarios should be covered, different keys to look up the \ac{CID} lengths may be beneficial.
In addition, it is up for discussion whether using \acp{CID} is beneficial or whether the five-tuple approach is sufficient.

\afblock{Flow Selection.}
Our prototype only measures flows that are explicitly added to a selection list.
More dynamic approaches are of interest for network operators, e.g., for dynamically suggesting flows to track.
Even though dynamically tracking flows is beyond the scope of this work, we prototyped a long header \ac{CID} extraction to validate the feasibility.

\hyphenation{di-rec-tion-al}

\afblock{Measurement Direction.}
We currently only perform one-directional measurements, i.e., track a flow passing in both directions as two separate flows.
The main reason for this choice is that we cannot guarantee that we can track a bi-directional flow in all cases as one pipe of the Tofino (similar to other switches) only manages a subset of all ports. 
If the ingress and egress ports of a bi-directional flow are on different pipes, the registers holding the flow information are not shared between those pipes, and thus, the information is inaccessible between the two.
Nevertheless, the control plane can map these flows.

\afblock{Time Resolution.}
As already discussed in \pref{sub:rtt_calculation}, our implementation uses a fixed 16-bit slice of the available 48-bit nanosecond resolution timestamp.
The lowest time interval supported by our implementation is $\approx$ 1.049 ms, and there is a rollover roughly every 69 seconds.
Our current prototype can only resolve one rollover between consecutive spin bit flips for which we invest additional hardware resources, i.e., we duplicate the \emph{Spin Bit Value Register} and \emph{Ring Buffer Register}.
Given typical communication intervals, we believe a rollover of 69 seconds to be a feasible solution.
However, given that Chen et al.~\cite{chen:spin2020:RTTmeasurements} use full nanosecond resolution, we plan to evaluate if the selected 16-bit slice is sufficient.

\afblock{Reordering Threshold.}
Our current prototype uses a reordering threshold of three.
While configurable, choosing a value is not straightforward as there might be, e.g., flows holding less than three packets per RTT.
Thus, similar to the square bit~\cite{ietf-draft-explicit-flow-measurements}, it is an open question how to configure the threshold in practice.

\afblock{Future Work.}
Our evaluation focusses on the general feasibility of our prototype, i.e., showing that pure spin bit measurements as well as additional post-processing and reorder protection can be implemented on Tofino. 
However, the measurement performance depends on several set screws which might further need tuning to the expected network settings.
We thus plan to perform an additional thorough evaluation of our prototype, e.g., inspecting the impact of different reordering thresholds, ring buffer sizes, or measurement read-out intervals, as well as different network conditions.

\section{Conclusion} 
\label{sec:conclusion}
The IETF's decision to encrypt almost all of QUIC's signaling information, including its notion of SEQs and ACKs, challenges the established network monitoring world as conventional solutions for tracking flow RTTs are no longer applicable.
While the spin bit addresses this challenge and allows for RTT measurements, it is susceptible to reorderings, which have to be filtered using additional methods of spin bit tracking software.
This, however, goes against the recent trend of increasingly performing networking monitoring tasks directly on programmable networking hardware.

This paper shows that the spin bit does not stand against this trend as it can be effectively tracked on an Intel Tofino switch.
Furthermore, simple mechanisms to protect against reordering prove to be quite effective and also map to the programmable data plane, as does additional post-processing of the measurements in the form of simple averaging and RTT-based classification.

However, some features of QUIC, such as not having a connection ID length indicator in short header packets, significantly complicate data plane implementations on the current generation of programmable hardware.

\begin{acks}
This work has been funded by the German Research Foundation DFG under Grant No. WE 2935/20-1 (LEGATO).
We thank the anonymous reviewers for their valuable comments.
\end{acks}


\clearpage
\balance
\bibliographystyle{ACM-Reference-Format}
\bibliography{literature}

\end{document}